\documentclass[final,3p,times,twocolumn,number]{elsarticle}

\usepackage{graphicx}

\usepackage[dvips]{hyperref}

\usepackage{amssymb}

\usepackage{amsthm}
\usepackage{textcomp}

\biboptions{sort&compress}

\begin{document}

\begin{frontmatter}

\ead{juergen.klepp@univie.ac.at}
\ead[url]{http://fun.univie.ac.at/}

\title{Neutron-optical gratings from nanoparticle-polymer composites}

\author[label1]{J. Klepp}
\author[label2]{C. Pruner}
\author[label3]{M. A. Ellabban}
\author[label4]{Y. Tomita}
\author[label5,label6]{H. Lemmel} 
\author[label5,label6]{H. Rauch}
\author[label1]{M. Fally}
\address[label1]{University of Vienna, Faculty of Physics, Boltzmanngasse 5, 1090 Vienna, Austria}
\address[label2]{University of Salzburg, Department of Materials Science and Physics, 5020 Salzburg, Austria}
\address[label3]{Tanta University, Faculty of Science, Physics Department, 31527 Tanta, Egypt}
\address[label4]{University of Electro-Communications, Department of Electronics Engineering, 1-5-1 Chofugaoka, Chofu, Tokyo 182, Japan}
\address[label5]{Vienna University of Technology, Atominstitut, Stadionallee 2, 1020 Wien, Austria} 
\address[label6]{Institut Laue Langevin, Bo\^{i}te Postale 156, F-38042 Grenoble Cedex 9, France}
\begin{abstract}
The preparation of neutron-optical phase gratings with light-optical holography is reviewed. We compare the relevant concepts of \emph{i)} Kogelnik's theory for Bragg diffraction of light by thick volume gratings, which can be used to analyze holographic gratings with both light and neutrons, and \emph{ii)} the dynamical theory of neutron diffraction. Without going into mathematical detail, we intend to illuminate their correspondence. The findings are illustrated by analyzing data obtained from reconstruction of nanoparticle holographic gratings with both light and neutrons.        
\end{abstract}


\end{frontmatter}

\section{Introduction}
\label{sec:Introduction}

The photorefractive effect, i.\,e. the light-induced change of the light refractive-index in nonlinear optical materials has been studied intensively since its discovery in 1966 \cite{AshkinAPL1966}. It was only in 1990 that it was realized by Rupp \emph{et al.} \cite{Rupp-prl90} that its analog for neutrons exists: The photo-neutronrefractive effect, i.\,e. the light-induced change of the neutron refractive-index of materials, has been investigated with cold neutrons in SANS experiments and used for neutron interferometry \cite{Schellhorn-phb97,Fally-apb02,Pruner-nima06,Fally-prl06}. Next to polymer-dispersed liquid crystals and deuterated polymethylmethacrylate a new material class seems promising for designing versatile neutron-optical devices such as beam-splitters and mirrors of high optical quality: nanoparticle-polymer composites. In comparison to holographic gratings recorded in the above-mentioned materials investigated earlier, the refractive-index modulation for neutrons of nanoparticle-polymer composites can be further adjusted by choosing the species of nanoparticles. Furthermore, these materials are resistant to volume shrinkage, which usually poses a problem in polymerization processes \cite{SuzukiAPL2002}. 

\begin{figure}
\begin{center}
\scalebox{0.4}
{\includegraphics {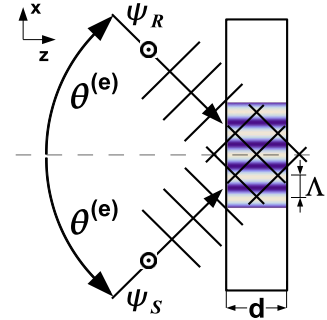}}
    \caption{Recording of holographic gratings with laser light. $d$ and $\Lambda$ are the mechanical thickness and spacing of the grating, respectively.}
    \label{fig1}
\end{center}    
\end{figure}

\section{Recording of holographic gratings with light}
\label{sec:Recording}

In Fig.\,\ref{fig1}, the experimental setup for recording holographic gratings in photosensitive materials is sketched. 
Two coherent plane s-polarized light waves of amplitude $A_0/2$ and wavelength $\lambda_L=2\pi/k_L$ in free space, whose $y$-component can be written as 
\begin{eqnarray*}
\psi_S&=&
A_0/2\exp\left[ik_L\left(z\cos\theta^{(e)}+x\sin\theta^{(e)}\right)\right],\\ 
\psi_R&=&
A_0/2\exp\left[ik_L\left(z\cos\theta^{(e)}-x\sin\theta^{(e)}\right)\right],
\end{eqnarray*}
are brought to interference within the material. Each beam encloses the (external) incidence angle $\theta^{(e)}$ with the sample surface normal. The resulting light pattern exhibits a sinusoidal modulation $I=|\psi_S+\psi_R|^2=A_0^2/2\left[1+\cos\left( K_Lx\right)\right]$, where $K_L=2\pi/\Lambda=2k_L\sin\theta^{(e)}$. 
For a given material, illumination time and intensity, the light-induced refractive-index change may be written as 
$\Delta n_L(x)=\Delta n_L^{(0)} + \Delta n_L^{(1)} \cos (K_Lx)$. 
In the present case, the photosensitive material consists of a nanoparticle-monomer-photoinitiator mixture \cite{SuzukiAPL2002,SuzukiApplOpt2004,TomitaOptLett2005}. The photoinitiator triggers polymerization upon illumination with light. In regions with high intensity, monomers are consumed by the polymerization reaction. More monomers diffuse from dark to bright regions. Therefore, nanoparticles (e.\,g. SiO$_2$) migrate to dark regions. 

\section{Diffraction from holographic gratings}
\label{sec:DiffrFromGratings}

The recorded holographic gratings can be used for diffraction of light or neutrons.  
In 1969, Kogelnik \cite{KogelnikBellSysTechJ1969} formulated a two-wave coupling theory for Bragg diffraction of monochromatic light by thick sinusoidal holographic gratings starting from the Helmholtz equation for the complex electric field amplitude inside the grating. The approach is valid for reflection and transmission holograms and also at large diffraction efficiencies $\eta_L=I_D/(I_D+I_F)$, with the intensity of the Bragg-diffracted beam $I_D$ and the intensity of the beam in forward direction $I_F$. It includes phase as well as absorption gratings. Moreover, also slanted geometries -- in contrast to the symmetric situation in Fig.\,\ref{fig1}, in which no refractive-index modulation occurs in $z$-direction -- can be treated easily.  
In particular, the theory assumes 
\begin{enumerate}
\item incidence near the Bragg angle,
\item only two coupled waves propagating inside the material,
\item slow energy variation in space of the two coupled waves compared to their wavelength (be it due to energy exchange or absorption), 
\item that second derivatives of wave amplitudes can be neglected (this is a consequence of the previous assumption),
\item that outgoing waves are equal to the waves inside the material. In other words, boundaries between vacuum or air (light refractive-index equal or close to 1) and the material (\emph{mean} light refractive-index larger than 1) are neglected. The mean refractive-index is assumed to be constant in space. Only the \emph{modulation} of the refractive-index $\Delta n^{(1)}_L$ is taken into account. This is again a consequence of the previous assumption, as is discussed in \cite{Gaylord82}.
\end{enumerate}
These assumptions lead to two first-order differential equations (coupled wave equations), either containing both wave amplitudes.

\subsection{Light} 
\label{subsec:Light} 

Solving the coupled wave equations mentioned above for the special case of lossless dielectric holographic unslanted gratings \cite{KogelnikBellSysTechJ1969}, the light diffraction efficiency reads as 
\begin{eqnarray}\label{eq:etaLight}
\eta_L=\nu_L^2
\mbox{sinc}^2\left(\!\sqrt{\nu_L^2+\xi_L^2}\right),
\end{eqnarray}
in transmission geometry (Laue case), with 
\begin{eqnarray*}
\nu_L=\kappaup\frac{d}{\cos\theta}=\frac{\pi\Delta n_L^{(1)} d}{\lambda_L\cos\theta},~\xi_L=\frac{\pi d\left(\vartheta_B-\theta\right)}{\Lambda}.
\end{eqnarray*}
Here, $\Delta n_L^{(1)}$, $\vartheta_B=\arcsin\left[\lambda_L/(2\Lambda n_L)\right]$ and $\theta=\arcsin\left(\sin\theta^{(e)}/n_L\right)$ are the amplitude of the light-induced sinusoidal refractive-index modulation for light, the Bragg angle and the actual angle of incidence (both as measured inside the material), respectively. $n_L$ is the mean refractive-index of the material for light including $\Delta n_L^{(0)}$. $\kappaup$ is the coupling constant giving the strength of coupling between the two waves that propagate through the material. $\Delta n_L^{(1)}$ arises from the modulation of the dielectric permittivity $\epsilon$ due to inhomogeneous distribution of nanoparticles and polymer.

\subsection{Neutrons} 
\label{subsec:Neutrons}

From the time-independent one-particle Schr\"{o}dinger equation (see e.\,g. \cite{Sears-89})
\begin{eqnarray*}
\left[-\frac{\hbar^2}{2m}\nabla^2+V(\vec r)\right]\psi(\vec r)=E\psi(\vec r),
\end{eqnarray*}
one obtains a Helmholtz-type wave equation 
\begin{eqnarray*}
\left[\nabla^2+K^2(\vec r)\right]\psi(\vec r)=0
\end{eqnarray*}
\cite{Rauch-00}, with $K(\vec r)=\sqrt{2m\left[E-V(\vec r)\right]}/\hbar$, the neutron wave number inside the material. $E=\hbar^2k^2/(2m)$ is the energy of the free neutron.
As usual, one defines the neutron refractive-index of the material as 
$n=K/k$, with the wave number $k=2\pi/\lambda$ of the incident neutron beam. The neutron-optical potential in nanoparticle-polymer holographic gratings is 
\begin{eqnarray}\label{eq:NeutronOptPot}
V=2\pi\hbar^2b_c\left[\rho +\Delta\rho(x)\right]/m,
\end{eqnarray}
with the neutron mass $m$, the coherent scattering length $b_c$ for a particular nanoparticle-polymer mixture and the number density $\rho$. The photo-neutronrefractive effect in nanoparticle-polymer composites arises from the modulation of the neutron-optical potential due to inhomogeneous distribution of nanoparticles resulting in $\Delta\rho(x)\ne 0$. The neutron refractive-index becomes 
\begin{eqnarray*}
n=\sqrt{1-\frac{V}{E}}\simeq 1-\frac{\lambda^2b_c\left[\rho +\Delta\rho(x)\right]}{2\pi}
\end{eqnarray*}
and therefore the neutron diffraction efficiency $\eta$ can be written in a similar form as in Eq.\,(\ref{eq:etaLight}), with $\nu_L$ and $\xi_L$ replaced by 
\begin{eqnarray}\label{eq:nuXiNeutrons}
\nu=\frac{\lambda\, d\, b_c\Delta\rho}{2\cos\theta},~\xi=\frac{\pi d\left(\theta_B-\theta\right)}{\Lambda},
\end{eqnarray} 
respectively. Here, the Bragg angle is $\theta_B=\arcsin\left[\lambda/(2\Lambda)\right]$. From the above equations, it is expected that the following parameters can be adjusted for fabrication of neutron-optical elements: 
First, the grating thickness $d$ can be freely chosen up to a material-dependent value at which extinction inhibits penetration of the material with light and therefore hologram recording. Second, the grating spacing $\Lambda$ is determined by $\theta^{(e)}$ for given $\lambda_L$. Unfortunately, $\Lambda_{min}=\lambda_L/2$ -- which is at least in the range of 100\,nm for laser wavelengths -- is the lower limit for given $\lambda_L$. This results in very small Bragg angles for neutrons. A possible solution to overcome this limit is triggering nonlinear processes in the material that generate higher harmonics of the refractive-index modulation, as has already been demonstrated in \cite{Havermeyer-prl98}. Third, $\Delta\rho$ can be adjusted by the illumination time and/or intensity. Finally, the coherent scattering length $b_c$ is determined by the particular choice of nanoparticles and their isotopes.

\section{Correspondence to the dynamical theory of neutron diffraction}

Dynamical diffraction theory for scattering of neutrons (see e.\,g. \cite{Rauch-78,Sears-89,Rauch-00}) in crystals is based upon solving the one-body Schr\"{o}dinger equation outside and inside the material, the latter being represented by a periodic optical potential.
It is assumed that the Born approximation is invalid, i.\,e. multiple scattering effects are taken into account. Furthermore, only two waves -- forward diffracted and Bragg-reflected -- are assumed to have considerable amplitudes. Employing a superposition of Bloch modes as ansatz and taking the boundary conditions into account, one finds two values for the refractive-index for each of the two internal waves and so the coherent wave is a superposition of four modes. One can calculate the reflectivity $R$ of a thick crystal in the symmetric Laue case as:
\begin{eqnarray}\label{eq:Reflectivity}
R(\mbox{x},\vary)=\frac{\sin^2\left(\vary\sqrt{\mbox{x}^2+1}\right)}{\mbox{x}^2+1},
\end{eqnarray}
with the dimensionless parameters 
\begin{eqnarray*}
\mbox{x}=\frac{\pi \mathbf{V}\sin(2\theta_B)}{\lambda^2|F|}\cdot(\theta-\theta_B)
,~\vary=\frac{\pi\,d}{\Delta}=\frac{\lambda\,d|F|}{\mathbf{V}\cos\theta_B}
\end{eqnarray*}
(x and $\vary$ should not be confused with the spatial coordinates $x$ and $y$ in Fig.\,\ref{fig1}).
Here, $\mathbf{V}$ and $|F|$ are the volume of the primitive unit-cell and the unit-cell structure factor for a particular reflection plane, respectively. The quantity $\Delta$ is called extinction length or Pendell\"{o}sung period. It determines the oscillation period of the internal amplitudes of forward diffracted and Bragg-reflected waves along the direction parallel to the surface normal ($z$-direction in Fig.\,\ref{fig1}). Pendell\"{o}sung interference fringes for Laue reflections from silicon crystals have been demonstrated by variation of the incident wavelength \cite{Shull-prl68}. The effect is also present in oscillations of the integrated reflectivity, that have been shown in \cite{Sippel-pl65}.
\begin{figure}
\scalebox{0.25}
{\includegraphics {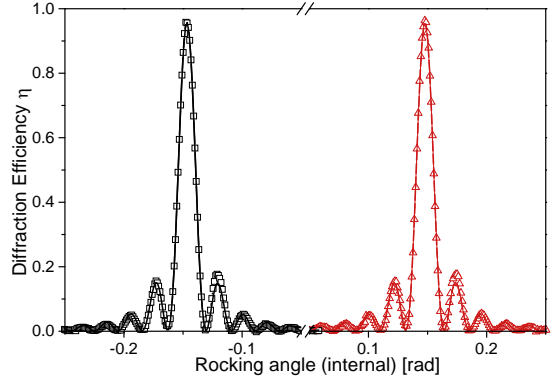}}
    \caption{$\pm$1$^{\mbox{\scriptsize{st}}}$ order rocking curves (left/right) of a nanoparticle holographic grating with 20 vol\% SiO$_2$ nanoparticles, $d\simeq50\,\mu$m and $\Lambda=1\,\mu$m measured with laser light of wavelength $\lambda_L=458\,$nm.}
    \label{fig2}
\end{figure}

Comparing Eqs.\,(\ref{eq:etaLight}) and (\ref{eq:Reflectivity}) one can see the following correspondence between the parameters $\nu,\,\xi$ used in the Kogelnik theory and $\vary,\,\mbox{x}$ used in the dynamical diffraction theory for neutrons: 
\begin{eqnarray*}
R&\leftrightarrow&\eta\\
\vary&\leftrightarrow&\nu\\
\mbox{x}&\leftrightarrow&\frac{\xi}{\nu}
\end{eqnarray*}
Calculating an analogue $\Delta_K$ to $\Delta$ via the relation $\nu=\pi\,d/\Delta_K$, one finds 
\begin{eqnarray}\label{eq:PendellPer}
\Delta_K=\frac{2\pi\cos\theta_B}{\lambda\,b_c\Delta\rho}.
\end{eqnarray}

\section{An example}
\label{sec:Example} 

Two glass microscope slides are glued together enclosing a fixed gap provided by a roughly $50\,\mu m$ thick spacer-foil. The liquid monomer-nanoparticle mixture is sucked into the gap by capillary force. After recording a hologram in the material as described in Sec.\,\ref{sec:Recording}, a laser beam of wavelength $\lambda_L=458$\,nm is diffracted by the holographic grating with 20 vol\% of SiO$_2$ nanoparticles of roughly 10\,nm diameter, grating thickness $d\simeq 50\,\mu$m and grating spacing $\Lambda=1\,\mu$m. By rotating the sample through the Bragg-positions for -1$^{\mbox{\scriptsize{st}}}$ and +1$^{\mbox{\scriptsize{st}}}$ diffraction orders, the rocking curves shown in Fig.\,\ref{fig2} are observed. A fit according to Eq.\,(\ref{eq:etaLight}) (solid lines) yields $d= 50.60(5)\,\mu$m, while the first-order refractive-index modulation for light is found to be $\Delta n_L^{(1)}= 5.007(9)\times 10^{-3}$. Absorption is negligible at this wavelength. 

\begin{figure}
\scalebox{0.37}
{\includegraphics {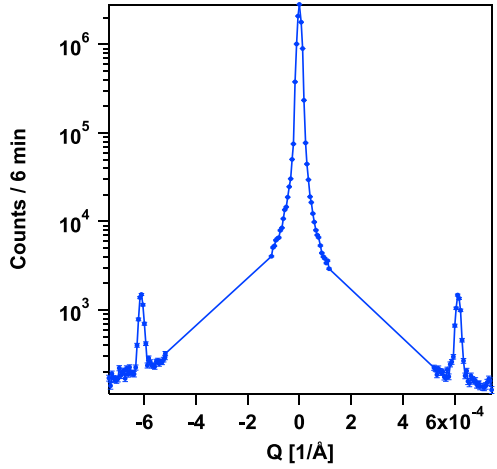}}
    \caption{Diffraction pattern of the same sample as analyzed in Fig.\,\ref{fig2}, measured by analyzer crystal rotation at a neutron wavelength of about $1.9\,$\AA.}
    \label{fig3}
\end{figure}
The sample has also been analyzed with thermal neutrons using the USANS-option of the interferometer instrument S\,18 of the Institute Laue Langevin (ILL), Grenoble \cite{KroupaNIMA2000}. This double perfect crystal diffractometer setup accesses scattering angles down to the $\mu$rad range, corresponding to $\mu$m structures in real space. The wavelength of the incident beam was $\lambda\simeq 1.9\,$\AA, roughly four orders of magnitudes less than the grating spacing of the sample. By rotation of the analyzer crystal at normal incidence of the beam to the sample, the diffraction pattern as shown in Fig.\,\ref{fig3} is measured. 
\begin{figure}
\scalebox{0.381}
{\includegraphics {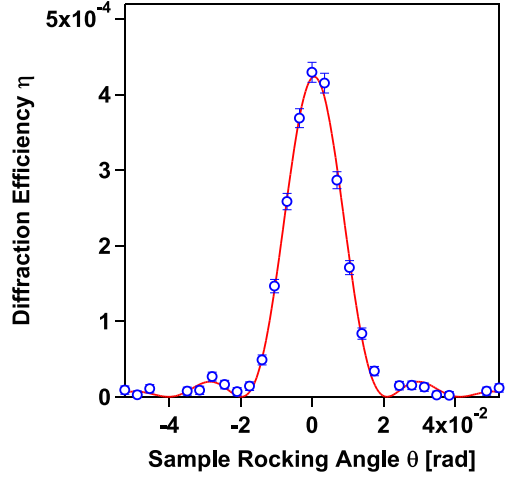}}
    \caption{Rocking curve of the same sample with analyzer crystal at -1$^{\mbox{\scriptsize{st}}}$ order peak position (c.\,f. Fig.\,\ref{fig3}).}
    \label{fig4}
\end{figure}
Fixing the analyzer crystal position at the -1$^{\mbox{\scriptsize{st}}}$ order peak (left peak in Fig.\,\ref{fig3}) and rotating the sample through the -1$^{\mbox{\scriptsize{st}}}$ order Bragg-angle, the rocking curve as shown in Fig.\,\ref{fig4} is observed. 
The solid line is a fit according to Eq.\,(\ref{eq:etaLight}) including Eqs.\,(\ref{eq:nuXiNeutrons}), which describes well the measured data. The fit yields the parameter values $d= 49.1(6)\,\mu$m and $b_c\Delta\rho= 4.36(5)\,\mu$m$^{-2}$. 
The Bragg angle at this wavelength is $\theta_B\simeq 6\times 10^{-3}$\,\textdegree. 
One can calculate the Pendell\"{o}sung period using Eq.\,(\ref{eq:PendellPer}) to find $\Delta_K\simeq 7\,$mm, a value that is about two to three orders of magnitude larger than for the (220) reflection of a silicon crystal slab. The Pendell\"{o}sung oscillations are described by
\begin{eqnarray*}
\!\!\!\!\!\!\!\!\!\!\!\eta(\theta_B)\!=\!\sin^2\!\left(\frac{\lambda b_c\Delta\rho}{2\cos\theta_B}\cdot d\right)\leftrightarrow
R(\theta_B)\!=\!\sin^2\!\left(\frac{\lambda |F|}{\mathbf{V}\cos\theta_B}\cdot d\right)
\end{eqnarray*} 
within the frame of the two corresponding theories compared in this paper.
From the above expressions it immediately becomes clear that this enormous value is mainly due to $b_c\Delta\rho$ being very small in holographic nanoparticle-polymer gratings.

\section{Discussion and concluding remarks} 

It is important to note that three diffraction orders are clearly visible in Fig.\,\ref{fig3}, so obviously three waves instead of two -- as was assumed in Sec.\,\ref{sec:DiffrFromGratings} -- are coupled propagating through the sample. Therefore, the grating can be considered to be a `thin' grating at this wavelength, i.\,e. diffraction occurs in the multi-wave coupling- or Raman-Nath-regime and \emph{not} in the two-wave coupling- or Bragg-regime (see e.\,g. \cite{GaylordApplOpt1982}). However, since the samples' diffraction efficiency for thermal neutrons is very small, the intensity that contributes to the additional (third) peak is negligible and two-wave coupling theory can be applied to analyze the measured rocking curve in Fig.\,\ref{fig4}. For larger values of $\eta$, rigorous coupled wave analysis \cite{MoharamJOptSocAm1981} must be employed if more than one reflected beam is observed.

It is well known that for thermal neutron diffraction by a thick silicon crystal in the Laue case, the reflectivity $R$ in Eq.\,\ref{eq:Reflectivity} oscillates rapidly as a function of the parameter x, that reflects the deviation from the Bragg-condition $\theta-\theta_B$ for a given crystal and wavelength. This is due to the high values of $\vary$, that are typically more than $10\,\pi$. As a consequence of limited resolution, the \emph{measured} reflectivity can be at most $\overline R=0.5$ at Bragg-angle. One of the major advantages of holographic gratings over perfect crystals is that such limitations can be avoided. The parameter $\nu$ -- corresponding to $\vary$ -- that determines the frequency of oscillation of $R$ depending on $\theta-\theta_B$ can be tuned by careful choice of $b_c$, $d$ and $\Delta\rho$ and is typically smaller than $\pi$. As a consequence, the diffraction efficiency $\eta$ or reflectivity $R$ can well exceed the limit of 0.5 and can, in principle, be equal to 1 at $\nu=\pi/2$. For instance, one can implement mirrors instead of the usual beam-splitters for the second grating of triple-Laue neutron interferometers. Used that way, mirrors save intensity that is lost in perfect-crystal neutron interferometry. Mirrors also prevent contribution to the background by parasitic beams whose signal cannot be separated from the signal of the interfering beams in the case of small Bragg-angles \cite{Pruner-nima06}.   

\section*{Acknowledgments}
This work was financed by the Austrian Science Fund FWF, Project nos. P 20265 and P 18988. The beam time at the ILL and 
financial support from the Ministry of Education, Culture, Sports,
Science and Technology of Japan under grant 20360028 is greatly acknowledged.

\bibliographystyle{model1a-num-names}
\bibliography{juergen}

\end{document}